\documentstyle[pra,aps]{revtex}

\newcommand{\BEL}{\begin{eqnarray}}
\newcommand{\EEL}{\end{eqnarray}}
\newcommand{\BE}{\begin{eqnarray*}}
\newcommand{\EE}{\end{eqnarray*}}
\newcommand{\DS}{\displaystyle}

\newcommand{\VF}[1]{\mbox{\boldmath$#1$}}
\newcommand{\VP}[1]{\mbox{$\vec{#1}$}}
\newcommand{\TP}[1]{\mbox{$\stackrel{\leftrightarrow}{\DS #1}$}}
\newcommand{\TF}[1]{\mbox{\boldmath$\stackrel{\leftrightarrow}{\DS #1}$}}

\newcommand{\kreuz}{\!\times\!}
\newcommand{\kr}{\kreuz}
\def\V#1{\VP{#1}}
\def\T#1{\TP{#1}}
\def\VecFett{\def\V##1{\VF{##1}}}

\def\TenFett{\def\T##1{\TF{##1}}}

{\begin{list}{}%
{\settowidth{\labelwidth}{\bf #1}%
\setlength{\leftmargin}{\labelwidth}%
\addtolength{\leftmargin}{\labelsep}%
}}%
{\end{list}}
\VecFett 
\TenFett
\begin{document}
\begin{center}
{\Large\bf Potential equations for plasmas round a rotating\\[0.5cm]
 black hole}\\[3cm]
{ Klaus Els\"asser }\\
Institut f\"ur Theoretische Physik\\
Ruhr-Universit\"at Bochum\\
D-44780 Bochum, Germany\\[5cm]
\end{center}

\begin{abstract}
The generalized Helmholtz equations of relativistic multifluid plasmas can be
integrated for axisymmetric equilibria in close analogy to the magnetic flux
conservation law in ideal magnetohydrodynamics (J. D. Bekenstein and E. Oron,
Phys. Rev. {\bf{D 18}}, 1809 (1978)). The results are, for each fluid
component, two flux functions and a potential equation for the poloidal stream
function. Amp\`ere's equation for the four-potential $A_\nu$ is reduced to two
coupled equations for the time-like and the toroidal component. So we have
altogether four potential equations for a two-component plasma; they can be
derived from a variational principle. 
\end{abstract}

\vfill
PACS: \hspace*{4ex}04. 40. Nr,\hspace*{4ex} 04. 70. Bw,\hspace*{4ex} 52. 60. +h,
 \hspace*{4ex}97. 60. Lf

\newpage
\section*{I. Introduction}

\hspace*{3ex}Observations of active galactic nuclei and other massive objects
have increased the interest in equilibrium plasma models within a Schwarzschild
or Kerr geometry. Even if the metric is given and all physical quantities are
assumed to be independent of a toroidal angle $\varphi$ and of time $t$, there is
still some effort needed to reduce the whole set of equations for the
electromagnetic field and the fluid quantities. This has been done within ideal
magnetohydrodynamics (MHD) by several authors with various degrees of
completeness and sophistication. A characteristic feature of the MHD description
is the conservation of magnetic flux in a system co-moving with the
center-of-mass four-velocity $u^\mu$ (of ions, essentially); this equation
(``Ohm's law'', Eq. (\ref{2}) below) has been ``integrated'' for general
stationary and axisymmetric systems by Bekenstein and Oron \cite{Bek78} who give
also a basic discussion and some historical background of general-relativistic
MHD. As a result, one can represent the electromagnetic field tensor
$F_{\mu\nu}$ completely by the particle flux $nu^\mu$, the two Killing vectors
associated with the translational symmetry in $t$ and $\varphi$, and two ``flux
functions'' which are constant along the poloidal stream lines (Eq. (\ref{17})
below). Further reductions of the whole set of MHD equations and discussions of
the astrophysical background have been given by Camenzind \cite{Cam86}, Mobarry and
Lovelace \cite{Mob86}, Nitta, Takahashi and Tomimatsu \cite{Nit91}, and Beskin
and Par'ev \cite{Bes93}. Thus one arrives, as in the non-relativistic case, at a
single potential equation for the magnetic flux function $\Psi$ (the covariant
toroidal component $A_\varphi$ of the vector potential $\V{A}$), together with
some constraints.

\hspace*{3ex}One problem with these MHD models is the large number of arbitrary
flux functions -- no dissipation mechanism has been included to reduce this
number --, and a fluid picture may be questionable if the collision frequencies
are too low. This, however, is not our point here, and we admit an ideal fluid
picture on reasons of simplicity. The question is, however, whether the usual
MHD theory is consistent for a really rotating black hole \cite{Kerr63}. In this case the
metric is necessarily non-diagonal; the relevant element $g_{t\varphi}$ is the
invariant scalar product of the two Killing vectors mentioned above
\cite{Strau84}, so it can by no means be transformed to zero. On the other hand,
it couples the components $A_t$ and $A_\varphi$ in Amp\`ere's law which
will, in general, contradict the strong coupling of $A_t$ and $A_\varphi$ in the
MHD models (here they have to be functions of each other). So we are led to the
question of how to reduce the set of multifluid plasma equations and Maxwell's
equations for stationary axisymmetric systems without the discrepancy mentioned
above. 

\hspace*{3ex}In Section II we give a short account of the work of Bekenstein and
Oron, leading to the strong coupling of $A_t$ and $A_\varphi$. The same
solution, however, can be used for any generalized Helmholtz equation in ideal
fluid descriptions. Two examples of these Helmholtz equations are derived in
Sections III and IV, respectively: One refers to the ordinary non-relativistic MHD equation (momentum
balance), the other to the relativistic multifluid plasma, assuming a constant
temperature of all species. The latter example is used for the reduction problem
in Section V. Here it is shown that the multifluid equations are reduced to one
single potential equation for each species, and the poloidal components of
Amp\`ere's equation are integrated.  A numerical
solution of this full set of four potential equations would be facilitated by the
existence of a variational principle; the solution could then be approximated
with finite elements by minimizing the corresponding functional. So it may be
interesting that such a functional exists, as is shown in Section VI. Finally we
discuss the results in Section VII.

\section*{II. Flux conservation in axisymmetric plasmas}

\hspace*{3ex}Let us assume a plasma configuration where all physical quantities,
including the metric tensor components $g_{\mu\nu}$, are independent of time $t$
and of a toroidal angle $\varphi$. We choose a coordinate system $(x^\mu)$ with
$x^0=ct \;(c=1)$, $x^1=\varphi$, and $x^2, x^3$ some poloidal coordinates. Assuming in
addition that all physical quantities are invariant to the simultaneous
inversion of $t$ and $\varphi$ -- which is reasonable for any rotating
equilibrium -- the most general line element $ds$ can be represented as follows
\cite{Cha92}:
\BEL
(ds)^2= g_{rs}\;dx^r\; dx^s + g_{ab}\; dx^a\;dx^b, \label{1}
\EEL
where the indices $r,s$ run from $0$ to $1$, and $a,b$ from $2$ to
$3$. We want to determine an electromagnetic field tensor $F_{\mu\nu}$ which is
consistent with Eq. (\ref{1}), and which, in addition, obeys the condition of
magnetic flux conservation in a medium moving with the Eulerian four-velocity
$u^\mu$. The latter condition is familiar from ideal magnetohydrodynamics (MHD)
and means that a certain four-vector $E_\mu$ (``co-moving electric field'')
should vanish:
\BEL
E_\mu\equiv F_{\mu\nu} u^\nu =0. \label{2}
\EEL
The field tensor $F_{\mu\nu}$, of course, should solve the homogeneous Maxwell
equations; therefore it can be written as the curl of a four-potential $A_\mu$
in the usual manner:
\BEL
F_{\mu\nu}= A_{\nu ,\mu}-A_{\mu ,\nu}, \label{3}
\EEL
where $( )_{,\mu}$ is the partial derivative of the quantity in brackets with
respect to $x^\mu$. Finally we use the continuity equation
\BEL
(\sqrt{-g}\;nu^\mu)_{,\mu}=0, \label{4}
\EEL
with $g$ the determinant of $g_{\mu\nu}$ and $n$ the particle number density in
the local inertial rest frame.

\hspace*{3ex}To solve Eq. (\ref{2}) we remember that any skew-symmetric tensor
$F_{\mu\nu}$ can be represented by two four-vectors, $E_\mu$ and $B_\mu$
(``co-moving magnetic field'') \cite{Bek78}; denoting all components in the
local inertial rest frame by primes (with $u^{0\prime}=c=1; u^{i\prime} =0$ for 
$i=1,2,3$) we
define
\BE
E^\prime_0 = 0\quad &;&\quad E^\prime_i=F^\prime_{i0},\quad i=1,2,3\\
B^\prime_0 =0\quad &;&\quad B^\prime_i=F^\prime_{jk},
\EE
where in the last equation $(i,j,k)$ is a cyclic permutation of $(1,2,3)$. In
this coordinate system we use the Minkowski metric
\BE
(g^\prime_{\mu\nu})=diag (1,-1,-1,-1),
\EE
and $E^{i\prime}=-E^\prime_i,\; B^{i\prime}=-B^\prime_i$ are the local electric and magnetic fields,
respectively. It is then easily seen that the covariant representation of $E_\mu$
in the laboratory frame with any velocity field $u^\mu(x)$ is given by the left
part of Eq. (\ref{2}), while $B_\mu$ is given by
\BEL
B_\mu = -{1\over 2}\varepsilon _{\mu\nu\varrho \sigma}u^\nu F^{\varrho \sigma},
\label{5}
\EEL
where the totally antisymmetric Levi-Civit\`a tensor $\varepsilon
_{\mu\nu\varrho \sigma}$ includes a factor $\sqrt{-g}$ in order to be a tensor.
Both vectors $E_\mu, B_\mu$ are obviously orthogonal to $u^\mu$; this
corresponds to $2\kr 3$ independent components. Therefore we can represent the six
independent elements of $F_{\mu\nu}$ by these vector components, namely:
\BEL
F_{\mu\nu}=E_\mu u_\nu -E_\nu u_\mu -\varepsilon _{\mu\nu\varrho
\sigma}u^\varrho B^\sigma. \label{6}
\EEL
The covariant equation (\ref{6}) can easily be proved by writing it in the local
inertial rest frame. Eq. (\ref{6}) is generally valid; in the case of flux
conservation according to Eq. (\ref{2}) it simplifies to
\BEL
F_{\mu\nu}=-\varepsilon _{\mu\nu\varrho \sigma}u^\varrho B^\sigma. \label{7}
\EEL
Obviously the field tensor $F_{\mu\nu}$ is then orthogonal not only to $u^\mu$,
as required by Eq. (\ref{2}), but also to $B^\mu$:
\BEL
B^\mu F_{\mu\nu}=0. \label{8}
\EEL
The remaining task is now to construct $B^\mu$ for axisymmetric equilibria.

\hspace*{3ex}For this purpose we consider the two Killing vectors $(\xi^\mu)$
associated with these symmetries, namely:
\BE
(\xi^\mu)=&(k^\mu)&\equiv (1,0,0,0),\\
(\xi^\mu)=&(m^\mu)&\equiv (0,1,0,0).
\EE
In both cases they lead to a vanishing partial derivative of any physical
quantity $A$ in the direction of $\xi^\mu$:
\BEL
\xi^\mu A_{,\mu}=0. \label{9}
\EEL
Let us assume that all components $A_\nu$ of the vector potential share this
property (though a gauge transformation could destroy it). Then Eq. (\ref{3})
leads to 
\BEL
F_{r\nu}\equiv \xi^\mu F_{\mu\nu}=-\xi^\mu A_{\mu ,\nu}=-A_{r,\nu},\label{10}
\EEL
where $r=0$ for $\xi^\mu = k^\mu$ and $r=1$ for $\xi^\mu =m^\mu$. The right-hand
side of this expression is obviously non-zero only for $\nu=a=2$ or $3$.
Differentiating Eq. (\ref{10}) with respect to $x^b$, $b=2$ or $3$, but $b\neq
a$, we find
\BEL
F_{ra,b}=-A_{r,ab}=F_{rb,a}.\label{11}
\EEL
To obtain the same expressions from Eq. (\ref{7}) we write the factor
$\sqrt{-g}$ of $\varepsilon _{\mu\nu\varrho \sigma}$ explicitely, with the
remaining constant permutation symbol $\tilde{\varepsilon }_{\mu\nu\varrho
\sigma}$:
\BE
\varepsilon _{\mu\nu\varrho \sigma}=\sqrt{-g}\;\tilde{\varepsilon }_{\mu\nu\varrho
\sigma}.
\EE
Then our integrability condition from Eqs. (\ref{7}) and (\ref{11}) reads as
follows:
\BE
0&=& F_{ra,b}-F_{rb,a}\\
&=& \tilde{\varepsilon
}_{rsab}\left[\left(\sqrt{-g}\;u^sB^b\right)_{,b}-\left(\sqrt{-g}\;u^b B^s\right)
_{,b}\right.\\
&& \left. \qquad\qquad +\left(\sqrt{-g}\;u^s B^a\right)_{,a}- \left(\sqrt{-g}\; u^a
B^s\right)_{,a}\right].
\EE
The right-hand side of this equation is understood with fixed and mutually
different values for $r,s,a$ and $b$. We can re-write it by restoring the
summation convention with respect to the index $a$:
\BEL
\left(\sqrt{-g}\; u^s B^a\right)_{,a}-\left(\sqrt{-g} \;u^a B^s\right)_{,a} =0;
\quad s=0,1. \label{12}
\EEL
The general solution $B^\mu$ of Eq. (\ref{12}) with $B^\mu u_\mu =0$ can be
written as follows:
\BE
B^\mu = \alpha  u^\mu + b^\mu \quad ;\quad \alpha \equiv -b^\mu u_\mu,
\EE
where $b^\mu$ solves the same Eq. (\ref{12}) as $B^\mu$. Since it is linear in
$b^\mu$ we can solve it separately for the poloidal part $b^a$ and for $b^s$.
Ignoring now $b^a$ we can solve Eq. (\ref{12}) for $b^s$ if the poloidal flow
$u^a$ is not identically zero. It is useful to represent $b^s$ by a linear
combination of the Killing vectors, namely:
\BEL
b^\mu &=& -n C\,K^\mu, \label{13}\\
K^\mu &\equiv & k^\mu + \beta m^\mu, \label{14}
\EEL
where the coefficients $C$ and $\beta$ have to be determined suitably. The factor
$n$ has been included in order to take advantage of the mass conservation law,
Eq. (\ref{4}), where the replacement $\mu\rightarrow a$ is allowed due to the
symmetries. Then we have a solution of Eq. (\ref{12}) if $C$ and $\beta$ are
constant along the poloidal stream lines (``flux functions''):
\BEL
u^a C_{,a} &=& 0\; = \; u^a \beta_{,a}, \label{15}\\
B^\mu &=& -n C \left[K^\mu -(K^\lambda  u_\lambda )u^\mu\right], \label{16}\\
F_{\mu\nu}&=& nC\varepsilon _{\mu\nu\varrho \sigma}u^\varrho K^\sigma
.\label{17}
\EEL
The last equation has been obtained by inserting Eq. (\ref{16}) into Eq.
(\ref{7}). While $C$ is an arbitrary flux function, $\beta$ is fixed by the
condition that $F_{\mu\nu}$ is perpendicular to both $u^\mu, B^\mu $ or $u^\mu,
K^\mu$. Using Eqs. (\ref{2}), (\ref{8}), (\ref{10}), (\ref{14}) and (\ref{16})
we find
\BEL
0= K^\mu F_{\mu\nu}=-(A_{0,\nu}+ \beta A_{1,\nu}). \label{18}
\EEL
This equation can only be fulfilled if $A_0$ and $A_1$ are functions of each
other and constant along the poloidal stream lines. This is indeed the case
as can now easily be shown from Eq. (\ref{2}) for $\mu =r$ and Eq.
(\ref{10}):
\BEL
0= F_{r\nu} u^\nu =-A_{r,a} u^a. \label{19}
\EEL
This completes our particular solution for $B^\mu$ and $F_{\mu\nu}$ if $b^a=0$.
It coincides with the result of Ref. \cite{Bek78} (their $A$ is our $- \beta$).
Why is a poloidal part of $b^\mu$ not possible? Our symmetry requires $F_{rs}=0$
according to Eq. (\ref{10}); from Eq. (\ref{7}) we have
\BE
F_{rs}= -\varepsilon _{rs\varrho \sigma}u^\varrho  B^\sigma = -\varepsilon
_{rsab} u^a B^b.
\EE
Here we see that for a non-zero poloidal flow $u^a$ the poloidal part of $B^\mu$
must be proportional to $u^\mu$, otherwise $F_{rs}$ would be non-zero. Therefore
the solution as given in Eqs. (\ref{16}) and (\ref{17}) is unique, up to the
specification of two flux functions, $C$ and $\beta$.

\hspace*{3ex}Sometimes it is useful to re-write the result of this section in
particular poloidal coordinates as defined by the poloidal stream lines, $\Psi$
= const., and an angle-like coordinate $\theta$ varying along the poloidal
stream lines:
\BE
x^2= \Psi\quad ;\quad x^3=\theta.
\EE
From Eq. (\ref{19}) we know that $A_r = A_r(\Psi)$, so we may identify one of
both components with $\Psi$ itself, while the other component defines $\beta$
according to Eq. (\ref{18}), e.g.:
\BEL
A_1 \equiv \Psi\quad ;\quad \beta = -dA_0(\Psi)/d\Psi.\label{20}
\EEL
In this coordinate system we have from Eq. (\ref{10}):
\BE
F_{13}=0\quad ; \quad F_{12} =-1.
\EE
Inserting here Eq. (\ref{17}) we find
\BEL
u^2=0\quad ;\quad C=1/(\sqrt{-g}\; nu^3).\label{21}
\EEL
The remaining elements of $F_{\mu\nu}$ in this coordinate system are then:
\BE
F_{0a}=-\beta F_{1a}\quad ;\quad F_{23}=(\beta u^0-u^1)/u^3.
\EE
Then our vector $B^\mu$ and field tensor elements $F_{\mu\nu}$ are, for a given
geometry and flow field, completely determined by one single flux function
$\beta$.
\section*{III. Vorticities in MHD plasmas}

\hspace*{3ex}Flux conservation laws are generally expected if dissipative
processes are negligible. Let us first discuss the non-relativistic case. The
oldest example is the Kelvin/Helmholtz theorem for a neutral fluid with pressure
$p=p(\varrho )$, where $\varrho $ is the mass density. Euler's equation may then
be written using the vorticity $\V{\omega }$ and the Bernoulli function $U$ as
follows:
\BEL
{\partial \V{v}\over \partial t}+\V{\omega }\kr \V{v}+\nabla U=0,\label{22}
\EEL
and we obtain immediately Helmholtz's equation for the vorticity by taking the
curl:
\BEL
{\partial \V{\omega }\over \partial t}+\nabla \kr (\V{\omega }\kr \V{v}) =0.
\label{23}
\EEL
This equation is the prototype of any vector field $\V{\omega }\rightarrow
\V{\Omega}$ which is ``frozen in'', moving with the fluid velocity $\V{v}$, and
which is the curl of another field, say $\V{V}$:
\BEL
{\partial \V{\Omega}\over \partial t}+\nabla \kr (\V{\Omega}\kr \V{v})=0 \; ; \;
\quad\V{\Omega}\equiv \nabla \kr \V{V}. \label{24}
\EEL
To get an equation for $\V{V}$ we integrate Eq. (\ref{24}), introducing a scalar
potential $\Phi$:
\BEL
{\partial \V{V}\over \partial t}+\V{\Omega}\kr \V{v}+\nabla \Phi =0. \label{25}
\EEL
For $\V{V} = \V{A}$ and $\V{\Omega} = \V{B}$ we obtain
\BEL
\V{E}+\V{v} \kr \V{B}=0, \label{26}
\EEL
where the electric field $\V{E}$ is the usual expression in terms of the
potentials $\V{A}, \Phi$ (with $c=1$). Eq. (\ref{26}) is just the non-relativistic limit of
Eq. (\ref{2}). For other ideal fluid models one may also find a Kelvin/Helmholtz
theorem, though the corresponding vectors $\V{V}, \V{\Omega}$ are more
complicated. The momentum balance of ideal MHD theory refers to a ``center-of-mass'' 
fluid with total pressure $p$, total mass density $\varrho $, center-of-mass 
velocity $\V{v}$, and the Lorentz force due to the electric current density
$\V{j}$:
\BE
\varrho {d\V{v}\over dt}+\nabla p=\V{j}\kr \V{B}.
\EE
The picture of an idealized fluid requires not only zero resistivity according
to Eq. (\ref{26}), but, more generally, zero entropy production. So we expect
for ideal MHD theory a second Kelvin/Helmholtz theorem and associated vectors
$\V{V}$, $\V{\Omega}$; they seem to be unknown, but they can be constructed
using appropriate Lagrangian and Lin variables \cite{Els94}, \cite{Spi98}. Here
we use three Lin variables: the entropy per mass $s$ and the two Euler
potentials $q_\lambda \;(\lambda =1,2)$ of the magnetic field:
\BE
\V{B}=\nabla q_1 \kr \nabla q_2.
\EE
The constraints of entropy and magnetic flux conservation are then expressed
by these three material invariants:
\BE
{ds \over dt}={dq_1\over dt}={dq_2\over dt} =0,
\EE
and the Ansatz for $\V{V}$ reads as follows:
\BEL
\V{V}=\V{v}-r\nabla s-\sum\limits^2_{\lambda =1}v_\lambda \nabla q_\lambda .
\label{27}
\EEL
The coefficients $r,v_\lambda $ are now determined in order to match Eq.
(\ref{25}) with an arbitrary potential $\tilde{\Phi}$. From Eq. (\ref{27}) and
its curl, the equation of $\V{\Omega}$, we find after some rearrangements the
following purely kinematical relation:
\BEL
{\partial \V{V}\over \partial t}+\V{\Omega}\kr \V{v}&=&{\partial \V{v}\over
\partial t}+\V{\omega}\kr \V{v}-{dr\over dt}\nabla s+{ds\over dt}\nabla
r\nonumber\\
&& -\sum\limits_\lambda {dv_\lambda \over dt}\nabla q_\lambda 
+\sum\limits_\lambda  {dq_\lambda \over dt}\nabla v_\lambda \nonumber\\
&& -\nabla \left(r{\partial s\over \partial t}+\sum\limits_\lambda v_\lambda
{\partial q_\lambda \over \partial t}\right).\label{28}
\EEL
Here we insert our particular fluid model for $\partial \V{v}/\partial t$. The
ideal MHD equation can be written as follows:
\BE
{\partial \V{v}\over \partial t}+\V{\omega}\kr\V{v}=-\nabla \left({\V{v}^2\over
2}+h\right)+T\nabla s+{\V{j}\over \varrho }\kr \V{B},
\EE
where $h$ is the enthalpy per mass, $T$ the temperature, and $\V{j}$ the current
density as determined from Amp\`ere's  law. The right-hand side of Eq.
(\ref{28}) is then just $-\nabla \tilde{\Phi}$, with
\BE
\tilde{\Phi}={\V{v}^2\over 2}+h+r {\partial s\over \partial
t}+\sum\limits_\lambda \, v_\lambda  {\partial q_\lambda \over \partial t},
\EE
provided the coefficients $r, v_\lambda $ obey the following equations of
motion:
\BE
{dr\over dt}=T\quad ;\quad {dv_1\over dt}={1\over \varrho }\;\V{j}
\cdot \nabla q_2\quad ;\quad {dv_2 \over dt}=-{1\over \varrho }\;\V{j}
\cdot \nabla q_1.
\EE
So we find a second flux conservation law in ideal MHD; it refers to the
following generalized vorticity:
\BEL
\V{\Omega}\equiv \nabla \kr \V{V}=\V{\omega}-\nabla r\kr \nabla s
-\sum\limits_\lambda \nabla  v_\lambda \kr \nabla q_\lambda .\label{29}
\EEL
The generalized Helmholtz equation (\ref{24}) allows, of course, the ``trivial''
solution $\V{\Omega}=0$, corresponding to a potential flow for $\V{V}$, $\V{V}=
-\nabla S$; then Eq. (\ref{27}) leads to the so-called Clebsch representation of
$\V{v}$:
\BE
\V{v}=-\nabla S+r\nabla s+\sum\limits^2_{\lambda =1} v_\lambda \nabla q_\lambda
.
\EE
(For $\V{\Omega}\neq 0$ we would need a further pair of Clebsch variables
$v_\lambda ,q_\lambda $ with $dv_\lambda /dt=dq_\lambda /dt=0$). In contrast to
the first Kelvin/Helmhotz theorem associated with Eq. (\ref{26}) we have no
simple advantage from this second vorticity law; in particular, the equations of
motion for $r$ and $v_\lambda $ have to be solved, in addition to the
conservation laws for $s$ and $q_\lambda $ and (not shown here) the Bernoulli
equation for $S$. The situation, however, becomes more transparent if we leave
the MHD description and treat the plasma as a fluid of different species $j
\;(j=e:$ electrons, $j=i:$ ions of any kind) interacting only via the
electromagnetic field, with the following momentum balance for each species:
\BE
\varrho _j\bigg({\partial \over \partial t}+\V{v}_j\cdot \nabla \bigg)\V{v}_j+\nabla
p_j=e_j n_j(\V{E}+\V{v}_j\kr \V{B}).
\EE
There is an obvious formal bridge to the MHD description for a two-component
plasma: Neglecting the electron mass ($m_e\sim \varrho _e\rightarrow 0$) we
obtain $\V{v}_i$ as the center-of-mass velocity $\V{v}$; assuming, in addition,
quasi-neutrality with simply charged ions ($n_e=n_i=n$) we can replace $\V{v}_e$
as follows:
\BE
\V{v}_e=\V{v}-{\V{j}\over |e|n}.
\EE
Adding both momentum equations gives then the correct ideal MHD balance with
$p=p_e+p_i$, while the momentum equation of electrons leads to a generalized
Ohm's law replacing Eq. (\ref{26}):
\BE
\V{E}+\V{v}\kr \V{B}={1\over |e|n}(\V{j}\kr \V{B}-\nabla p_e).
\EE
This ``modified'' MHD theory is the bridge mentioned above, but a simpler form
of the two Kelvin/Helmholtz theorems is obtained if we come back to the
multfluid description, now within general relativity.
\section*{IV. Vorticities in isothermal plasmas}

\hspace*{3ex}The flux conservation theorems associated with perfect fluid models
depend crucially on the equation of state for the pressure, e.g., $p=p(\varrho ,s)$.
The non-relativistic case in standard textbooks on neutral fluids usually
assumes constant entropy $s$ throughout the fluid volume, leading to the usual
Helmholtz equation, Eq. (\ref{23}). The corresponding equation in general
relativity has been derived by Taub \cite{Taub59}. If the entropy of a
non-relativistic neutral fluid varies in space, the situation is more subtle; it
has been discussed already by Eckart \cite{Eck60} and later in Ref.
\cite{Els94}. The result is that the vector $\V{V}$ whose vorticity flux is
conserved differs from $\V{v}$ by $(-r\nabla s)$, the second term on the
right-hand side of Eq. (\ref{27}). For a non-relativistic multifluid plasma Eq.
(\ref{27}) is replaced by
\BE
\V{V}_j=\V{v}_j-r_j\nabla s_j+{e_j\over m_j}\V{A},
\EE
leading to a ``three-circulation theorem'' corresponding to the three
constituents of $\V{V}\;(\V{A}$ is the vector potential). In general relativity we
have, instead of Eq. (\ref{25}) (from which the general Helmholtz equation
(\ref{24}) follows), the equation corresponding to Eqs. (\ref{2}) and (\ref{3})
for each species (suppressing now the species index $j$):
\BE
u^\nu\Omega_{\mu\nu}&=&0,\\
\Omega_{\mu\nu}&\equiv &V_{\nu,\mu}-V_{\mu,\nu}.
\EE
The vector $V_\nu$ for a multifluid plasma with $s=$ const. has been given by
Lichn\'erowicz \cite{Lich67} and Carter \cite{Car79} (``single constituent
perfect fluid'', his example (c) in \S 4), and for varying $s$ by
Ref. \cite{Els97}, namely:
\BE
V_\mu={\sigma\over \varrho }u_\mu +rs_{,\mu}+{e\over m}A_\mu,
\EE
where $\sigma$ is the relativistic enthalpy per volume, and $u_\mu$ the
covariant Eulerian four-velocity (with $c=1)$. One of Carter's results refers
also to a neutral fluid with two constituents and varying entropy, but the
resulting canonical momentum per volume ($=\varrho V_\mu$) reads, in our notation,
as follows (from Eqs. (4.25), (4.26), (4.32) and (4.33) of Ref. \cite{Car79}):
\BE
\varrho V_\mu\equiv n\pi_\mu= \sigma u_\mu,
\EE
so the term $rs_{,\mu}$ is missing.

\hspace*{3ex}Here we consider a different physical situation which may be of astrophysical
interest: We assume that a radiation field acts like a heat reservoir for
electrons and protons, maintaining a constant temperature of them. Then the
term $rs_{,\mu}$  is again  absent, and we find the
relativistic Helmholtz equation for the ordinary canonical vorticity of both
species. In this case, $\sigma$ will turn out to be the free relativistic 
enthalpy per volume.
We start with the material energy-momentum-stress tensor
$T_\mu\;^\nu$ of an ideal electron or ion fluid without denoting the species index
explicitely. Since both fluids are only coupled via the electromagnetic field by
the Coulomb/Lorentz force, we can write the energy-momentum balance for both
species as follows:
\BEL
T_{\mu}\;{^\nu} _{;\nu}= en \; F_{\mu\nu}u^\nu.\label{30}
\EEL
The semi-colon indicates the covariant derivative. The coupling of both species
by the right-hand side of Eq. (\ref{30}) implies now that the magnetic flux is
not conserved, neither for the electron nor for the ion fluid. The tensor
$T_\mu\;^\nu$ for an ideal fluid is well-known (see, e.g., Ref. \cite{Wein72}):
\BE
T_\mu\;^\nu =\sigma u_\mu u^\nu -p\;\delta^\nu_\mu.
\EE
The last term in this equation is the scalar pressure $p$ in the local inertial
rest frame times the Kronecker symbol, and the scalar $\sigma$ depends on the
equation of state. Using also the mass conservation, Eq. (\ref{4}), we obtain
\BE
T_{\mu}\;{^\nu} _{;\nu} =\varrho u^\nu \left({\sigma\over \varrho }
u_\mu\right)_{;\nu}-p_{,\mu}.
\EE
This four-vector must be orthogonal to $u^\mu$, as is also the right-hand side
of Eq. (\ref{30}). With the normalization of $u^\mu \;(u_\mu u^\mu=1$) we find
then
\BE
0&=& u^\mu T_{\mu}\;{^\nu} _{;\nu}=\varrho u^\nu \left({\sigma\over \varrho
}\right)_{,\nu}-u^\nu p_{,\nu},\\
&& \quad d\left({\sigma\over \varrho }\right)={1\over \varrho }dp,
\EE
where the differential $d$ means variation along the path of a fluid element. In
the co-moving inertial rest frame we use the Gibbs-Duhem relation for the free
enthalpy $\mu$ per mass:
\BEL
d\mu = -s\;dT+{1\over \varrho }dp. \label{31}
\EEL
Ignoring temperature variations, and including the relativistic rest energy
$\varrho c^2$ (with $c\neq 1$ for the moment) the resulting expression for
$\sigma$ is then as follows:
\BEL
\sigma =\varrho \left(1+{\mu\over c^2}\right). \label{32}
\EEL
To derive a flux conservation law we insert the vector potential for
$F_{\mu\nu}$ in Eq. (\ref{30}) and put all terms to the left-hand side:
\BE
u^\nu \left[\left({\sigma \over \varrho }u_\mu \right)_{;\nu}+{e\over m}
A_{\mu;\nu}-{e\over m}A_{\nu;\mu}\right]-{1\over \varrho }p_{,\mu}=0.
\EE
To eliminate the pressure term we calculate the partial derivative from Eqs. (\ref{31}) and (\ref{32}) with
$dT=0$ (and this time $c=1$):
\BE
{1\over \varrho }p_{,\mu}=\left({\sigma\over \varrho }\right)_{,\mu}=u^\nu
\left( {\sigma\over \varrho }u_\nu\right)_{;\mu},
\EE
where in the last step we used again the normalization of $u^\nu$. This is now
just the term leading to flux conservation for the canonical vorticity of each
species; we define
\BEL
V_\mu &\equiv & {\sigma \over \varrho }u_\mu + {e\over m}A_\mu,\label{33}\\
\Omega_{\mu\nu} &\equiv & V_{\nu ;\mu}-V_{\mu ;\nu}=V_{\nu ,\mu}-V_{\mu ,\nu},
\label{34}
\EEL
and we find 
\BEL
\Omega_{\mu\nu} u^\nu =0.\label{35}
\EEL
The solution of Eq. (\ref{35}) for axisymmetric equilibria is now simply obtained from the previous section:
We replace there $F_{\mu\nu}$ by $\Omega_{\mu\nu}$ and $A_\mu$ by $V_\mu$. The
mean velocities of electrons and ions, however, are usually different; the
fluxes of their general vorticities $\Omega_{\mu\nu}$ are therefore conserved in
different frames. It is interesting to rewrite Eq. (\ref{35}) for the spatial
components of the canonical velocity, $V^i$, in the special-relativistic case
(no gravity), namely:
\BE
{\partial \V{V}\over \partial t}+\V{\Omega}\kr \V{v}+\nabla U=0,
\EE
where $U(\equiv cV_0)$ is the relativistic Bernoulli function, and $\Omega^i\;
(\equiv -\Omega_{jk})$ are the spatial vector components associated with
$\Omega_{\mu\nu}$. This equation is now again of the same form as Eq.
(\ref{25}), and we recover Eq. (\ref{24}) as the prototype of any
special-relativistic generalized Helmholtz equation. 
\section*{V. Multifluid plasma equations for axisymmetric equilibria}

\hspace*{3ex}Let us use a general poloidal coordinate system to solve the mass conservation
law, Eq. (\ref{4}), for each species separately by introducing an appropriate
stream function $\chi$, namely $(\tilde{\varepsilon }^{1ab}$ is again the
permutation symbol, here for spatial indices):
\BEL
u^a ={1\over \sqrt{-g}\;n}\tilde{\varepsilon }^{1ab}\chi_{,b}.\label{36}
\EEL
The poloidal stream lines are then given by $\chi=$ const., and Helmholtz'
equation (\ref{35}) in the symmetry plane, $\mu=r=0,1$, reads as follows:
\BE
0=-V_{r,\nu} u^\nu =-V_{r,a} u^a.
\EE
Similarly as in Eq. (\ref{19}) we conclude that the components $V_r$ can be any
flux functions with respect to $\chi$:
\BE
V_r =V_r (\chi).
\EE
These two flux functions and the corresponding components of $A_\nu$ fix two
components of $u_\nu$ for each species up to a factor $\varrho /\sigma$, namely:
\BEL
u_r={\varrho \over \sigma}\left(V_r(\chi)-{e\over m}A_r\right).\label{37}
\EEL
Assuming that $\chi$ and $A_\nu$ are given elsewhere we may read the
normalization condition for $u_\nu$ as an equation for $n$:
\BEL
1=g^{rs}u_r u_s +g_{ab} u^a u^b . \label{38}
\EEL
So all fluid quantities besides $\chi$ are determined by Eqs. (\ref{36}) -
(\ref{38}), and we find all elements of $\Omega_{\mu\nu}$ except $\Omega_{ab}$:
\BEL
\Omega_{rs}=0\quad ;\quad \Omega_{ra}=-V_{r,a}=-V^\prime_r \chi_{,a} ,\label{39}
\EEL
where the prime means differentiation with respect to $\chi$. The general
solution for $\Omega_{\mu\nu}$, however, can be obtained from Eqs. (\ref{17})
and (\ref{18}) with appropriate changes of notation, namely:
\BEL
\Omega_{\mu\nu}&=& nC(\chi)\varepsilon _{\mu\nu\varrho \sigma} u^\varrho
K^\sigma,\label{40}\\
\beta &=&-V^\prime_0(\chi)/V^\prime_1(\chi). \label{41}
\EEL
Comparing this with the results above (Eqs. (\ref{36}) and (\ref{39}) ) we find
the flux function $C(\chi)$:
\BEL
C(\chi)=-V^\prime_1(\chi). \label{42}
\EEL
The equation for the stream function $\chi$ is then obtained from Eq. (\ref{40})
with $\mu=2$ and $\nu=3$, where the left-hand side is determined according to
the definitions (\ref{34}) and (\ref{33}); the result is then the following:
\BEL
\Omega_{23}&=& \sqrt{-g}\; nV^\prime_r(\chi)u^r, \label{43}\\
\Omega_{23}&\equiv & \left({\sigma\over \varrho }u_3\right)_{,2}-\left({\sigma \over
\varrho }u_2\right)_{,3} +{e\over m}F_{23}.\label{44}
\EEL
The complete set of fluid equations (\ref{36}) - (\ref{38}) and (\ref{43}) -
(\ref{44}) for the unknown variables $u_r$, $n$ and $\chi$ is written for any
poloidal coordinate system, thus allowing any number of particle species. To
solve finally Amp\`ere's equation in the poloidal plane we are then free to use
particular coordinates and a particular gauge of $A_\nu$. To be consistent with
the usual notation we denote the projections of the $j^\mu$-lines onto the
poloidal plane the lines $\Psi =$ const., where the flux function $\Psi$ may
be identified with a ``radial'' coordinate $x^2$ as in Sec. II, and the stream
function $\sim I$  of $j^a$ is denoted as a flux function, $I=I(\Psi)$. The
continuity equation for $j^\mu$ in the poloidal plane is then solved as follows:
\BEL
j^2=0\quad ;\quad 4\pi \sqrt{-g}\; j^3 =I^\prime(\Psi), \label{45}
\EEL
where the prime of $I$ means differentiation with respect to $\Psi$. The flux
function $I(\Psi)$ is, of course, not independent from the stream functions
$\chi$ of the different species. From
\BE
j^\mu=\sum\limits_j enu^\mu
\EE
and Eq. (\ref{45}) we find the following relation:
\BEL
\sum\limits_j e\chi = -{I(\Psi)\over 4\pi}+ const., \label{46}
\EEL
where the sum over $j$ is the sum with respect to the different particle
species. A convenient gauge of $A_\nu$ is, as in the non-relativistic case, the
condition that $\V{A}$ is tangential to the surfaces $\Psi$=const. of the
current lines:
\BE
\V{A}\cdot \nabla \Psi=0.
\EE
This equation can usually be fulfilled by an appropriate gauge function since
$\Psi$ is a radial coordinate; for $\Psi=x^2$ it reads
\BEL
A_2 =0 \quad\mbox{or}\quad F_{23}=A_{3,2}.\label{47}
\EEL
Let us now start with Amp\`ere's equation for $A_\nu$:
\BEL
\left[\sqrt{-g}\; g^{\mu\varrho }g^{\nu\sigma}\left(A_{\sigma,\varrho
}-A_{\varrho ,\sigma}\right)\right]_{,\nu}=-4\pi \sqrt{-g}\; j^\mu. \label{48}
\EEL
In the symmetry plane, $\mu=r=0,1$, this equation is decoupled from the poloidal
components of $A_\nu$; inserting the fluid quantities for $j^r$ we have then the
following set of two equations for the components $A_s, s=0,1$:
\BEL
\left[\sqrt{-g} g^{rs} g^{ab}A_{s,b}\right]_{,a} = 4 \pi \sqrt{-g}\;g^{rs}\sum\limits_j
e \;n{\varrho \over \sigma}\left(V_s(\chi)-{e\over m}A_s\right).\label{49}
\EEL
In the poloidal plane, Eq. (\ref{48}) can be re-written with indices $a,b,c,d$
which run from 2 to 3 only:
\BE
\left[\sqrt{-g}\, g^{ab} g^{cd}F_{bd}\right]_{,c}=-4\pi \sqrt{-g}\, j^a
\EE
The left-hand side is easily evaluated due to the antisymmetry of $F_{bd}$; we
introduce the determinants of $g_{ab}$ and $g_{rs}$ explicitely:
\BE
g_{pol}\equiv det(g_{ab})\; ;\; g_{sym}=det(g_{rs}),
\EE
then we obtain the following form of Amp\`ere's equation in the poloidal plane
for $a=2$:
\BEL
\left[\sqrt{-{g_{sym}\over g_{pol}}}F_{23}\right]_{,3}=-4\pi \sqrt{-g}\; j^2,
\label{50}
\EEL
and for $a=3$:
\BEL
-\left[\sqrt{-{g_{sym}\over g_{pol}}}F_{23}\right]_{,2}=-4\pi \sqrt{-g}\; j^3.
\label{51}
\EEL
Inserting here Eq. (\ref{45}) we identify the expression in brackets as a flux
function, namely:
\BEL
\sqrt{-{g_{sym}\over g_{pol}}}F_{23}=I(\Psi). \label{52}
\EEL
In the non-relativistic case this is simply the covariant toroidal component of
the magnetic field. Finally, the component $A_3$ is determined from Eqs.
(\ref{52}) and (\ref{47}); it is not a flux function, and it is not needed in
the remaining set of equations.
\section*{VI. A variational principle}
\hspace*{3ex}The numerical solution of our potential equations for $\chi$, Eqs.
(\ref{43}) - (\ref{44}), and $A_r$, Eq. (\ref{49}), is simplified by the
existence of a functional of $\chi$ and $A_r$ which is stationary in
equilibrium. We need, however, still another quantity whose variation leads to
the normalization condition, Eq. (\ref{38}). So we look for a functional $W$ of
three quantities, $W=W (\sigma, \chi, A_r)$ say, whose independent variations
lead to the equilibrium conditions. The particle density $n$ (which is not
varied), and $\sigma$ of each species are then obtained afterwards from a
combination of the normalization condition an an equation of state according to
Eq. (\ref{32}).

\hspace*{3ex}Let us start with the normalization condition, where $u^a$ and 
$u_r$ are given by Eqs. (\ref{36}) and (\ref{37}), respectively.
The latter equation gives
\BEL
\delta u_r =-{1\over \sigma} u_r \; \delta \sigma +{\varrho \over
\sigma}V^\prime _r (\chi)\delta \chi -{en\over \sigma}\delta A_r, \label{63}
\EEL
while $\delta u^a$ depends on $\delta \chi$ only:
\BEL
\delta u^a= {1 \over \sqrt{-g}\;n}\tilde{\varepsilon}^{1ab}(\delta
\chi)_{,b}.\label{64}
\EEL
It is then easily realized that $W$ could be of the following form:
\BE
W(\sigma, \chi, A_r)=\int d^2 x\; \sqrt{-g}\sum\limits_j{\sigma\over
2}(g_{ab}u^au^b -g^{rs} u_r u_s -1)+\cdots,
\EE
where the terms indicated by dots are independent of $\sigma$, and the
integration is done in a fixed region of the poloidal plane $(x^2, x^3)$. It is
a remarkable effect that in the expression above the usual invariant $u_\nu
u^\nu$ is replaced by $u_a u^a - u_r u^r$ which is only invariant under
transformations in the poloidal plane. A similar replacement will be needed in
the invariant which produces Maxwell's equations in vacuum. Using our solution
for $F_{23}$, Eq. (\ref{52}), and $F_{rs}=0, F_{ar}=A_{r,a}$, we find 
\BE
-{1\over 16\pi}F_{\mu\nu}F^{\mu\nu}=-{1\over 8\pi}{I^2(\Psi)\over (-g_{sym})} -
{1\over 8\pi} g^{rs}g^{ab}A_{r,a}A_{s,b}.
\EE
Changing now the sign of the last expression above, we are led to the following
functional:
\BEL
W(\sigma,\chi,A_r)=\int d^2x\sqrt{-g}&\Bigg[&\sum\limits_j {\sigma \over
2}(g_{ab}u^a u^b - g^{rs}u_r u_s -1)\nonumber \\
 &-& \left.{1\over 8\pi}{I^2(\Psi)\over (-g_{sym})}+{1\over 8\pi}g^{rs}
g^{ab}A_{r,a}A_{s,b}\right]\label{65}
\EEL
Variations with respect to $\chi$ and $A_r$ are now done by eliminating
derivatives of $\delta \chi$ and $\delta A_r$ by partial integrations, assuming
that $\delta\chi$ and $\delta A_r$ vanish at the boundary. Furthermore, we have
to vary $I(\Psi)$ according to Eq. (\ref{46}), namely:
\BEL
\delta I (\Psi)=(-4\pi e)\delta \chi. \label{66}
\EEL
The total variation of $W(\sigma,\chi, A_r)$ is then obtained with the following
result:
\BEL
\delta W=\int d^2x\sqrt{-g}&&\left\{\sum\limits_j{1\over 2}(g_{ab}u^a u^b
+g^{rs}u_r u_s -1)\delta\sigma\right.\nonumber\\
&&+\sum\limits_j{1\over (-g_{sym})}\left(\tilde{\Delta}\chi +eI(\Psi)+g_{sym} 
\varrho u^rV^\prime_r(\chi)\right)\delta\chi\nonumber\\
&&+\left.\left[\sum\limits_j enu^r-{1\over 4\pi\sqrt{-g}}(\sqrt{-g}g^{rs}g^{ab}A_{s,b}
)_{,a}\right]\delta A_r\right\},\label{67}
\EEL
\BEL
\tilde{\Delta}\chi &\equiv &\sqrt{-{g_{sym}\over
g_{pol}}}\sum\limits_{a,b}(\tilde{g}_{ab}\chi_{,b}-\tilde{g}_{bb}\chi_{,a})_{,a},
\label{68}\\
\tilde{g}_{ab} &\equiv &{\sigma\over \sqrt{-g}\;n^2}g_{ab}.\label{69}
\EEL
The vanishing factor of $\delta A_r$ is easily identified with Amp\`ere's
equation in the symmetry plane, Eq. (\ref{49}). To identify the vanishing factor
of $\delta \chi$ we note that from Eqs. (\ref{44}), (\ref{52}), and (\ref{36})
we have
\BE
m\Omega_{23}=\sum\limits_{a,b}(\tilde{g}_{ab}\chi_{,b}-\tilde{g}_{bb}\chi_{,a})_{,a}
+e\sqrt{-{g_{pol}\over g_{sym}}}I(\Psi).
\EE
Inserting this in Eq. (\ref{43}) with a slight rearrangement, we find
\BEL
\tilde{\Delta}\chi +eI(\Psi)+g_{sym}\varrho u^r V^\prime_r(\chi)=0.\label{70}
\EEL
This is just the condition that $W$ is stationary with respect to variation of
$\chi$, assuming that $g_{sym}$ is finite at this point.
\section*{VII. Discussion}

\hspace*{3ex}A plasma equilibrium near a rotating black hole has been considered
in the ideal fluid picture. The usual MHD equations imply two flux conservation
laws. One is the well-known conservation law of magnetic flux (Ohm's law with
vanishing resistivity); it leads to two constants of motion for stationary
axisymmetric systems: The covariant time-like and toroidal components, $A_t$ and
$A_\varphi$, of the vector potential are constant on the poloidal stream lines of
the plasma bulk velocity. This well-known fact is in contradiction with
the coupling of $A_t$ and $A_\varphi$ for a metric with $g_{t\varphi} \neq 0$,
as is appropriate for a rotating black hole. Simple Grad-Shafranov type MHD
equilibria (see, e. g., Ref. \cite{Haz92} are then ruled out in this case. 
The second flux conservation law of
MHD has been derived in Section III for the non-relativistic case (Eqs.
(\ref{24}) and (\ref{27}) - (\ref{29})); it is, however, not as simple as the
magnetic flux conservation law. A more reasonable description for $g_{t\varphi}
\neq 0$ is given by the multifluid equations: They can be cast into the form of
Helmholtz equations for each fluid component \cite{Els97}; they are particularly
simple for isothermal plasmas, as is shown here (Eqs. (\ref{33}) - (\ref{35})).
Since the form of these equations is exactly the same as the magnetic flux
conservation law, we obtain two constants of motion for each species of an
axisymmetric equilibrium, the time-like and toroidal components of the canonical
velocity $V_\nu$. The component $V_t$ is the relativistic Bernoulli function,
and $V_\varphi$ is the canonical angular momentum per mass of a fluid particle
of a particular species; this has to be expected for axisymmetric equilibria
if the fluid components interact only through the electromagnetic field. The
poloidal components of Amp\`ere's equation can be integrated, too, by adjusting
the coordinates to the lines of the electric current, $\Psi=$ const.; the
relevant flux function $I(\Psi)$, Eq. (\ref{52}), corresponding to the toroidal
magnetic field, is now simply related to the stream functions $\chi$ of the
different plasma components according to Eq. (\ref{46}). The whole set of
potential equations can now be summarized in a functional $W(\sigma,\chi,A_r)$,
where $\sigma$ is the free enthalpy density of a particular species, $\chi$ the
stream function of its poloidal velocity, and $A_r$ stands for $A_t$ and
$A_\varphi$. The total variation of $W$ produces then the nomalization condition
for the Eulerian four-velocity of each species, the potential equation for
$\chi$ (Eq. (\ref{70})), and Amp\`ere's equations for $A_t$ and $A_\varphi$. 

\hspace*{3ex}It is interesting to consider possible solutions of these equations
in a given geometry with $g_{t\varphi} \neq 0$ like Kerr's metric. Equilibria
with poloidal velocity fields are of interest from an observational point of
view, because they are able to exchange mass, angular momentum etc. between
inner and outer parts. Plasma models with pure poloidal velocity fields,
however, are not possible; the reason is that the toroidal velocity $u_\varphi$ is proportional to $(V_\varphi
-(e/m) A_\varphi)$ (Eq. (\ref{37})), where $V_\varphi$ is constant on the stream
lines of the particular species which is considered, but $A_\varphi$ generally
not. Equilibria with pure toroidal velocities are possible for an
electron-positron plasma. In this case Eq. (\ref{70}) is solved trivially with
$\chi\equiv 0$, $V_r\equiv $ const., $V^\prime_r\equiv 0$, and the velocity components
$u_t$, $u_\varphi$ become equal but opposite in sign, up to a constant $V_r$
; they are determined from
Amp\`ere's equation (\ref{49}) which is highly nonlinear due to the
normalization condition. This solution, however, seems to be artificial, and an acceptable solution will exhibit 
inevitably also poloidal velocity components.

\newpage

\end{document}